\documentclass{ws-procs975x65}

\newcommand{\ds}{\displaystyle}
\newcommand{\q}{\,\,\,}

\begin{document}

\title{Chaos in the Yang-Mills theory and cosmology: quantum aspects}

\author{Sergei Matinyan}

\address{Yerevan Physics Institute, Alikhanian Brs.St. 2, Yerevan 375036, Armenia\\
ICRANet, Piazzale della Repubblica 10, 65100 Pescara, Italy\\ 
$^*$E-mail: smatinian@nc.rr.com}

\begin{abstract}
I describe the footprints of the classical chaos of the Yang-Mills fields in the quantum description.
I also review the behavior of the BKL chaotic approach to the classical singularity on the basis of 
the Loop Quantum Gravity.
\end{abstract}


\bodymatter

\section{Introduction: Classical chaos in the Yang-Mills field theory and General 
Relativity}
 
At the MG IX I had a talk on the chaos of the non-abelian gauge theory and the gravity (mainly 
cosmology). This talk was devoted to the classical fields. They are the gauge theories and are 
non-linear. Thus, chaoticity is not surprising generally,although we know examples of the stable 
solutions (solitons and others) to the non-linear equations. Concerning the Yang-Mills (YM) theory 
it is known that this theory is not integrable. All attempts to prove its integrability, not 
given up until now, are not conclusive and if one analysis them he will be convinced that the 
proof contains some conjecture (mostly of the mathematical nature).Of course some approximation scheme 
(e.g. large $N$-expansion) leads to the integrability.

Therefore, it is not surprising that in 1981 it was found that the classical sourceless YM
equations exhibit the strong chaotic phenomena. This fact was proved initially for the simplest form 
of the spatially homogeneous YM equations for $n= 2,3$ numbers of degrees of freedom [1,2] and then 
was extended to the spherically symmetric field theory [3].
The spatial-temporal chaos of YM fields was established by the lattice calculations 
(see book [4] for review of this activity). This gave rise to a program to describe the
YM dynamics not in terms of the potential and fields but rather in terms of the loops
(strings) variables [5].

These results demonstrate that the classical YM fields lack any special stable configuration, all 
states are chaotic and no particular configuration dominates in the Minkowski space-time in contrast 
to the Euclidean case, where the instantons give a dominant contribution to the functional integral.

Turning now to the problem of the chaos in the General Relativity (GR) we have to remark that 
despite an even longer history of the study the chaos in GR, the chaoticity 
is observed mostly in the problem of the approaching to the space like singularity $t = 0$.
These studies are initiated and developing now due to the famous BKL chapter of GR
[6]. The problem of the chaos in GR closely related to the problem of the singularity
in the classical GR, where it is unavoidable according to the fundamental singularity
theorems [7]. The backward evolution of an expanding universe leads to a singular state where the 
classical theory fails to be applied.

If, as remarked by Landau (see [8]) one considers the metric $g$ as a function of the synchronous 
time $t$ only, at some finite time interval $g=det(g_{\mu\nu})$ tends to zero as $t\rightarrow 0$,
independent of the equation of state or the character of the gravitational field, and this 
results the singularity.

The corresponding metric $ds^2=dt^2-dl^2$ $(\alpha, \beta =1,2,3)$, $dl^2= \gamma_{\alpha\beta} 
dx^{\alpha}dx^{\beta}$ near the singularity $t = 0$ is
\begin{equation}
dl^2=t^{2p_1} dx_1^2 + t^{2p_2} dx_2^2 + t^{2p_3} dx_3^2	           %(1)
\end{equation}
with 
$$
\sum_{i=1}^{3}p_i=1=\sum_{i=1}^{3} p_i^2.
$$
 
This is so-called Kasner solution [9] corresponding to the Bianchi I spatial geometry. Thus one 
arrives to the regular flat, homogeneous, anisotropic space with the total volume homogeneously 
approaching the singularity.
BKL consider the Bianchi IX geometry which generalize (1) and corresponds to the 
diagonal homogeneous anisotropic spatial metric with $\gamma_{II}=a_I^2$
\begin{equation}
dl^2=a_I^2 n_{\alpha}^In_{\beta}^I dx^{\alpha} dx^{\beta}           	%(2)
\end{equation}
with the unit vectors along the axes $n_{\alpha}^I$, $(I = 1, 2, 3)$. $a_I$  positive scale factors 
are functions of $t$ only.

The diagonal form of the matrix is a result of the vacuum Einstein equation $R_{0\alpha} = 0$ .
However, even in the presence of matter, as argued by BKL, the possible non-diagonal 
terms do not affect the character of the Kasner epochs and the character of the "replacement" of the 
Kasner exponents. As a result, the evolution towards singularity 
proceeds via a series of successive oscillations during which the distances along two of
the principal axes oscillate while they shrink monotonically along the third axis (Kasner
epochs). The volume $V = \int \sqrt det \gamma dx_1 dx_2 dx_3 = 16 \pi ^2 a_I a_{II} a_{III} 
\propto t$ as $t\rightarrow 0$.

A new "era" begins when the monotonically falling metric components begin to oscillate while one 
of the previously oscillating directions begin to contract. This approach to the singularity reveals 
itself as an infinite succession of alternating Kasner "epochs" and has the character of a random 
process [10].
Thus infinite number of the oscillations are confined between any finite time t and moment of $t = 0$.

The central role in the BKL approach to the singularity plays the justification that the time 
derivatives dominate over space ones at the approach to the singularity.
This fact allows to think that the inhomogeneous model can be well described by 
Bianchi IX model where the spatial geometry can be viewed as an assembly of the small patches 
each of which evolves almost independently. In other words, the dynamical decoupling of the different 
spatial points on the space-like slices has place. Sizes of the patches are defined by the scale of 
the space derivatives during the evolution while the curvatures grow. To sustain the homogeneity of 
the evolution, parches have to be subdivide more and more at the vicinity of the singularity. If the 
geometry by some reason is discrete, such fragmentation must stop [11].

In the recent paper [12] this view was justified by the generalization the Misner's Mixmaster model 
to the generic inhomogeneous case. It is shown that neglect of spatial 
gradients possible in the asymptotic regime. In other words, authors of [12] claim that 
the generic cosmological solution near the singularity is isomorphic, point by point in space, to 
the one of the Bianchi VIII and IX models because the spatial coordinates in the Mixmaster model enter 
as parameters, and one can apply the long wavelength approximation.

The global chaoticity of the classical YM fields and the chaotic behavior in the approach to the 
singularity in the classical GR, of course, have to be changed when the quantum effects enter the game.

For the YM fields we have to take into account the quantum effects since QCD
describes the quantum world of the hadrons and their interactions.
For distances close to the Planck scale the gravitational field acquires large curvature and the 
evolution to the $t=0$ singularity has to be replaced by the quantum dynamics.

In this situation, we expect that the chaos of the classical fields should be diminished if not eliminated 
completely due to the quantum fluctuations of the gauge and quark fields. Below we show how it is happened 
and to which extent. We will see in general that the chaotic phenomena of the YM fields do not disappear 
in full, exhibiting the explicit footprints of the classical chaos in the quantum world.
 
\section{Quantum chaos of the YM fields}
 
The quantum insight into the YM dynamics clearly is achieved if we consider the spatially homogeneous 
potentials $A^a_{\mu}(t)$ $(a = 1,2,3; \mu = 0,1,2,3)$ (YM classical and quantum mechanics).

It is obvious that if this type of fields exhibits classical chaos, as we know, the wider class of the 
non-homogeneous YM fields $A^a_{\mu}(x, t)$ also will have this property. We mentioned that in the 
Introduction. 

The YM sourceless equation for potential $A^a_{\mu}(t)$ in the gauge $A^a_0(t) = 0$ is 
reduced to the discrete Hamiltonian system (see  [4] for review)
\begin{equation}
d^2A_i^a/dt^2+g^2(A^a_j A^b_j A^b_i -  A^b_j A^b_j A^a_i)       		%(3)
\end{equation}
With the conserved "external" and "internal" angular momenta
\begin{equation}
M_i=\epsilon_{ijk} A^a_j \dot{A^a_k},													      %(4)
\end{equation}
\begin{equation}
N^a=\epsilon^{abc} A^b_i \dot{A^c_i},													      %(5)
\end{equation}
which are vanished for the sourceless fields. Thus there exist seven integrals of motion 
and system is not integrable.

Homogeneous limit of YM equations, or their long wavelength regime corresponds 
to the gluon high density $n_g$ and/or strong coupling regime $g^2 n_g \lambda >> 1$. In this 
sense it is stated that the homogeneous fields are the relevant degrees of freedom for infrared 
regime [13]. Very recently, the equations (3) were obtained as a strong coupling limit of YM 
fields [14]. It was shown that in the leading order of $1/g$ YM equations are reduced to (3). 
The resulting theory is stable and leads to the mass gap with the confinement. Author even 
calculated in this approach the glueball spectrum which is in good agreement with the lattice 
QCD computations.

Consider now the simplest case of two and three degrees of freedom $n=2,3$ with $x = A^1_1, 
y = A^2_2, z = A^3_3$. We obtain the systems of two (three) coupled oscillators with the potential 
energies 
\begin{equation}
V(x,y)=(g^2/2) \, x^2 y^2,													                %(6)
\end{equation}
\begin{equation}
V(x,y,z)=(g^2/2)\,\, (x^2y^2+y^2z^2+x^2z^2).                        %(7)
\end{equation}
Classically these systems, despite their extremely simple forms, exhibit strong chaotic behavior. 
Potential (6) ($x^2 y^2$-model) has been used in various fields of science, including chemistry, 
astronomy, astrophysics and cosmology (chaotic inflation).
We mainly describe here the case $n = 2$.

Quantum mechanical system with the potential (6) has only discrete spectrum [15] despite its 
open hyperbolic channels along the axes and its infinite phase space. Physically it is clear why 
it is so: quantum fluctuations, e.g. zero modes forbid the trajectory to escape along the axis 
where the potential energy vanishes. The system is thus confined to a finite volume and this 
implies the discreteness of the energy levels. Classically of course the "particle" always can 
escape along one of the axes without increasing its energy.

Despite so drastic influence of the quantization on the behavior of the system (6) chaos 
left its footprints: periodic (unstable) orbits of the classical potential (6) after quantization 
have so called scars [16] (See also [17,18]). Energy level spacing distribution for the system (6) has 
the Wigner-Dyson type distribution in contrast to the Poissonian one for the systems whose classical 
counterparts are regular. They are in accordance with the Random Matrix Theory for GOE.

One can go further and propose that the traces of the classical chaos, in principle,
should show in the real spectra of hadrons. For instance, if we would collect the rich enough 
glueball spectra then their mass spacing distribution has to reflect the chaoticity
of the classical gluon field (gluon statistics has to be deal with the assembly of the particles 
with the same quantum numbers) [19]. Not having today (when?) such a rich collection of glueballs 
as a cleanest sample, author of [20] used the relatively rich baryon and meson spectra for the 
examine the nearest-neighbor level spacing distributions for mass $m < 2.5$ GeV. It is seen that 
these distributions are well described by the Wigner surmise corresponding to the statistics of the GOE.

Of course, one should consider this result as very preliminary since barions and mesons with their 
quark content are not the best case to check this idea (see [4] to inquire why the glueballs are 
necessary to solve this problem).

We would like to stress the important role which play here the so called billiards 
(classical and quantum). If we write the potential $V(x^2 y^2)$ in the form $(x^2 y^2)^{1/\alpha}$
where $0 \leq \alpha \leq 1$, then limit $\alpha = 0$ corresponds to the so-called hyperbolic 
billiards, where the classical trajectory undergoes elastic collisions on an infinite barrier 
(hyperbolic cylinder $x^2 y^2= 1$). Trajectories lie in the $x-y$ plane and consist of rectilinear 
segments constructed by the rules of geometrical optics. The notion of billiards plays an important 
role also in the GR [21].

Consider now the quantum mechanical adventures of the coupled YM oscillators
in the study of the partition function for our non-integrable system (6)
\begin{equation}
Z(t)=T_r\,\left[exp\,(-t\hat{H})\right]=\sum^{\infty}_{n=0}\,e^{-tE_n}       %( 8 )
\end{equation}
with the quantum Hamiltonian $\hat{H}$ 
\begin{equation}                                                             %( 9 )
\hat{H}=\frac{\hbar^2}{2}\,\left(\frac{\partial^{2}}{\partial x^2}+\frac{\partial^{2}}{\partial 
y^2}\,\right)+\frac{g^2}{2} \,x^2\,y^2
\end{equation}
using the quite effective method of the adiabatic separation of the motion in the hyperbola 
channels of the equipotential curves $xy = const$ [22].

The partition function defines the integrated density states $N(E)$ by the inverse 
Laplace transform of $Z(t)$ 
\begin{equation}
N(E)=\int^E_0\,dE^{'}\,\rho (E^{'})=L^{-1}\,\left(\frac{Z(t)}{t}\right)      %( 10 )
\end{equation}
and for the large enough energy levels $E$ is given by the Thomas-Fermi term - 
the zero order term of the Wigner-Kirkwood expansion [23] 
\begin{equation}                                                             %( 11 )  
Z_0(t)=\frac{1}{\sqrt{2\pi}\,g\hbar^2\,t^{3/2}}\,\left(ln\,\frac{1}{g^2\hbar^4\,t^3}+9ln2-C\right]
\end{equation}
with $C$ the Euler constant.

From (11) and (8) one obtains $N(E) \sim E^{3/2} log E$. For the hyperbola billiard $(\alpha = 0)$
the computations give [24], with the logarithmic precision, $N(E) = \frac{1}{8\pi}\,E\,log E$.

This result some time ago was encouraging from the point of view of the famous 
Hilbert-Polya-Berry program to look for the quantum (classically chaotic) Hamiltonian
whose eigenvalues reproduce the Riemann zeta-function's zeros. However, from 
Random Matrix Theory we know that such Hamiltonian (or some operator ) must be
non-invariant with respect to time inversion (GUE not GOE!).

Returning to the calculation of $Z(t)$, we apply the method of [22] (see also [25]):
The range of the integration over $x$ and $y$ variables and momenta $p_x$ and $p_y$ in the 
calculation of $Z(t)$ in the Winger representation is divided into two regions:  
the central region ($/x /,/y / \leq Q$ ,where Q is arbitrary, not specified scale) and the 
channels region ($Q \leq /x/, Q \leq /y / $).In the central region it is natural to apply 
Wigner-Kirkwood method; in the channels with the "slow" motion in the x variables (along the channel) 
again will be used the Wigner-Kirkwood expansion, but the "fast" motion, transverse to the channel, 
must be treated quantum mechanically. Remarkably, the dependence on $Q$ from both regions is cancelled 
in the leading terms $(t Q^4 )^{-1} << 1$ up to eighth order of $\hbar$. No doubts are left that the 
higher order of $h$ terms behave similarly (see [25]). Thus, only $Q$-independent, non-leading terms 
are contributing to $Z(t)$.

We encounter here the phenomenon of the "transmutation" of small parameters:
Classical parameter $1/t\,Q^4$ which rules the adiabatic separation of the variables in the channels 
transmutes into the small quantum parameter of the final asymptotic series for $Z(t)$.

It is interesting to follow the motion in the channel (along the $x$ axis) a little in 
detail. Motion in the channel can be describe by the Hamiltonian 
\begin{equation}
H_y = \frac{1}{2}\,P^2_y + \frac{1}{2}\,\omega^2_x\,y^2                    %( 12 )
\end{equation}
where $\omega=g\,x$ is $x$ dependent frequency and eigenvalues of (12) are 
$(n+1/2)\,h\,g\,x$.

In the channel $(/x/>>/y/)$ where the derivatives w .r. t. $x$ are small relative to the
derivatives w. r .t. $y$, we may first average the motion over the quantum fluctuations 
of $y$ [25,26] described by (12) and by the corresponding wave function involving
Hermite polynomials with the frequency $\omega=g\,x$. The corresponding average 
$$
< n / H_y / n >= (n + \frac{1}{2} ½)\,\hbar\,g\,x 
$$
then appears as an effective potential for the motion in the "slow" variable $x$ 
\begin{equation}                                  %( 13 ) .
\left( -\frac{\hbar^2}{2}\,\frac{\partial^2}{\partial x^2}+(n+1/2)\,g\,\hbar\,x\right)\, 
\Phi_n\,(x) = E\,\Phi_n\,(x).
\end{equation}

This is the well known Schrodinger equation for a linear potential having the solutions in 
terms of Airy functions which shows the linear confinement and discrete spectrum for eigenvalues.

We would like to emphasize that this confinement is not like standard phenomenon commonly 
refered to as quark confinement. Here the potential described by the gauge field amplitude $x(t)$. 
One may call this phenomenon as "self-confinement":
the fields themselves "prepare" the effective potential barrier, prohibiting escape to the infinity.

As we remarked above, just due to this consistent treatment of the motion in the channels when 
each $n$-th quantum evolves along the $x$ axis according the Hamiltonian
\begin{equation}                                  %( 14 )
H^{(n)}_x = \frac{1}{2}\,P_x^2 + (n+1/2)\,\hbar\,g|x|
\end{equation}
there are the precise cancellation of all leading quantum corrections (in the regime 
$\frac{1}{t\,Q^4}<<1$) and only non-leading but $Q$-independent corrections survive and 
lead to the final answer for $Z(t)$ in the form of asymptotic series with the expansion 
parameter $g^2h^4t^2$.

One remark is worthy on the discreteness of the spectrum. In the corresponding supersymmetric 
quantum mechanics, due to the cancellation between the bosonic and fermionic modes, 
there appears the continuous spectrum which coexists with the discrete one [27] and the 
confinement generally has not place.

There is another approach to calculate $Z(t)$: add to the potential (6) the extra term 
$V^2(x^2+y^2)$ - Higgs vacuum term, compute $Z$ and then put $V = 0$ [28]. This limit 
opens the hyperbola channels classically. In quantum mechanics, this limit leads to the 
singularities:
Logarithmic for the Thomas-Fermi term and power like for the higher quantum corrections 
$V$ for the $k$-th order of $h$. However, quantum mechanics cures that: it introduce the 
Higgs-like term [29] to the potential and due to this the limit $V=0$ in $Z(T)$ has no 
singularity at all.

In conclusion of this Section we state that the lessons derived from this quantum mechanical 
study of the higher order corrections to the homogeneous limit of YM equations would be useful 
for the better understanding the internal dynamics of the YM quantum field theory.
 
\section{BKL chaos in the Loop Quantum Gravity (Loop Quantum Cosmology)}
 
In this Section we consider which kind modifications on should expect for the oscillating 
chaotic approach to the classical singularity $t=0$ (BKL scenario) if one includes the 
quantum corrections. It is natural that at each novel scheme of the quantum effects to the 
gravitational field, BKL chapter is tested with the various conclusion and verdicts about 
this scenario. I described this efforts briefly in the Talk to MG IX mentioned in the 
Introduction [30]. Now the number of these considerations of the BKL behavior is increased 
significantly and includes Matrix models, String theory, its brane aspects, higher derivative 
corrections, matter content modifications (dilaton, p-form fields) etc. (see e.g., [31]). In 
the most of these approaches classical singularity, as a rule, are not avoided and this has a 
strong influence on the BKL scheme.

Here we describe only one approach based on the Loop Quantum Gravity (LQG) [32] and its 
sibling, Loop Quantum Cosmology (LQC) which is based not only on the LQC, but includes 
several additional assumptions and simplifications [33].

Here we describe very briefly the basic notions of LQG and LQC.

As it is known, this approach to the canonical Hamiltonian G R is based on the use not the 
ADM variables (spatial metric and extrinsic curvature) but the Ashtekar variables [34,35] 
based on the inclusion the spin connection variables what allows the formulation closely to 
the YM-like gauge field theory.

Briefly, the following has a place.
In the expression of the contra variant spatial metric $g^{ab} = e_i^a\, e_i^b$ in terms of 
triad (orthogonal and normalized at each point) there is redundancy due to an arbitrary 
3-dimensional rotation of the triad which does not change metric.

Densitized triad 
\begin{equation}                      
E_i^a = g^{1/2}\, e_i^a,\,\,\,\,\,\, (g = det\,g_{ab})                %(15)
\end{equation}
together with the SU(2) connection $A^i_a$ (Ashtekar connection), j = 1, 2, 3; a, b =
1, 2, 3) form the pair ($A_a^i$, $E_i^a$) canonically connected to the metric conjugated 
pair ($g_{ab}$, $p^{ab}$). As in the gauge theories, "momentum" $E_i^a$ of the spin 
connection $A_a^i$ is an analog of the electric field of SU(2) YM theory (index labels 
a basic element of the SU(2) Lie algebra). Connections $A_a^i$ involve the curvature of 
space and spin connection $\Gamma_a^i$
\begin{equation}                         
A_a^i = \Gamma_a^i +\beta\, K_a^i                                   %(16) 
\end{equation}
where $K_a^i$ ("torsion") defines the extrinsic curvature $K_{ai} = K_{ab}\,e_i^b$.
Positive parameter $\beta$ was introduced by Barbero [36] as substitute of the former imaginary 
unit, to have a certain reality conditions for Ashtekar variables.

This $\beta$-ambiguity leads to the Poisson bracket dependent on $\beta$: 
\begin{equation}
\{A^i_a(x),\,E^b_j(y)\} = 
\kappa\,\beta\,\delta^b_a\,\delta^i_j\,\delta\,(x, y) \,\,\,\,\,\,
(\kappa=8\pi\,G)                                                      %(17)
\end{equation}
and after the Dirac first class constraint quantization, leads to the $\beta$-dependent 
Hamiltonian constraint which rules the evolution of the system ADM-like way.

The rest two constraint: Gauss constraint (generating triad rotations) and 
diffeomorphism constraint (generating spatial diffeomorphisms) are independent 
on the Barbero-Immirzi ambiguity.

We would like to remark that parameter $\beta$ can be considered as the rescaling 
of the triad
\begin{equation}                                  %( 18 ) 
E^a_i = \frac{1}{\beta}\,\sqrt{g}\,e^a_i
\end{equation}
and this allows the possibility to introduce the conformal symmetry [37].

However, in the Dirac quantization this will lead to the new first class 
constraint corresponding to the conformal symmetry.

The main advantage of the new variables is that they allow a natural 
smearing of the basis fields ($A$, $E$) to the linear objects without 
introducing a background and retain the well-defined algebra. The 
connections integrated along a curve, exponentiated with a path-ordered 
way. Thus we arrive to the holonomies (this is analogous of the 
quantum mechanics where the Heisenberg operator, e.g., $x$ is 
represented by Weyl operator $e^{ix}$):
\begin{equation}
\begin{array}{c}\ds
h_e\,[A] = P\,exp\,\int_e\,\tau_i\,A^i_a\,\dot{e}^a\,dt \,\,\,\,\, 
(\cdot = \frac{d}{dt}) \\[4mm]\ds                                 %( 19 ) 
\tau_i = -\frac{1}{2}\,i\,\sigma_i,
\end{array} 
\end{equation}
$\sigma_i$ are the Pauli matrices.

Trace of $h_e\,[A]$ corresponds to the Wilson loop for the closed curve in the YM 
theory. Similarly, we arrive to the fluxes by integrating densitized triad over 
two-surfaces $S$: 
\begin{equation}
F_s\,[E] = \int_s\,\tau^i\,E^a_i\,n_a\,d^2x                       %(20)
\end{equation}
where $n_a$ is the normal to the surface $S$.

The above introduced smearing, without introducing background, eliminate all 
delta functions in the Poisson relations giving the well-defined algebra 
to construct the Hilbert space. We do not dwell on the fundamental problem 
of LQG of this construction in terms of the holonomies and fluxes in the 
conditions of the diffeomorphism invariance. There are wide spectrum of the 
conceptual and the technical problems for LQG (see e.g. [38]).

We only make some remarks:

\begin{list}{}
\item * Barbieri- Immirzi ambiguity can be resolved "experimentally" comparing 
the LQG results with the Bekenstein-Hawking entropy for the Black Hole [39].
Taking the formula of the area eigenvalues $A=8\pi\beta l_p^2 \sqrt{j(j+1)}$ one obtains  $\beta=ln2/\pi \sqrt3$. The new data yield 0.27398.
\item * It was remarked [40] that the role of parameter $\beta$ in the canonical quantum 
gravity is analogous in various senses to that of the parameter describing 
the different sectors associated with the topological structure of the finite gauge
transformations in the YM theory. In contrast to the LQG, the $\theta$-term enters
as the Pontryagin topological term which is a total derivative.
\item * After quantization, homologies and fluxes act as well defined operators, 
fluxes have the discrete spectra. Since the spatial geometry is determined by the 
densitized triads, spatial geometry is discrete as well,with the discrete area
and volume operators. In this sense, it is sometime declared that the Quantum 
Gravity is a Natural Regulator of matter [41].
By this reason, differential equation for the Wheeler-De Witt constraint is 
replaced by the difference equations.
\item * Although in the LQG there is some parallel with the Wilson loops of YM 
theory (hence the term Loop Quantum Gravity), there is essential difference. 
In the YM theory different sizes of loops are inequivalent in the light of the
interpretation relying to the quark confinement (e.g., inside the large loops 
fields are chaotic, inside the small loops are regular, small-large 
w.r.t . the confinement radius). The value of the Wilson loop is invariant 
under continuum deformations only for the vanishing field strength.
In the LQG, with its diffeomorphism invariance, there is no physical 
information in the shape and size since two networks of the different shape 
but the same topology can always be related by a suitable diffeomorphism,
independently of the "value" of the Ashtekar field strength.
\end{list}

After these remarks we turn to the BKL problem for Bianchi IX space geometry. 
We will base here on the mini-superspace spanned by the scale factors $a_I$ 
of (2), diagonal anisotropic model, $g_{II} = a^2_I\,(t)$.

Hamiltonian constraint is written in terms of the spin connections $\Gamma_I$
(below we follow [42]).
%( 21 )
\begin{equation}
\begin{array}{c}\ds
	H=\frac{2}{\kappa}\,[(\Gamma_{J}\Gamma_{K}-\Gamma_{I})\,a_{I}-\frac{1}{4}\,a_{I}\dot{a}_{J}\dot{a}_{K}
  +(\Gamma_{I}\Gamma_{J}-\Gamma_{K})\,a_{K}\\[4mm]\ds
  -\frac{1}{4}\,a_{K}\dot{a}_{I}\dot{a}_{J}
  +(\Gamma_{K}\Gamma_{I}-\Gamma_{J})\,a_{J}-\frac{1}{4}\,a_{J}\dot{a}_{K}\dot{a}_{I}]\, \end{array}
\end{equation}

%( 22 )
\begin{equation}
	\Gamma=\frac{1}{2}\,\left(\frac{a_J}{a_K}+\frac{a_K}{a_J}-\frac{a^{2}_{I}}{a_J\,a_K}\right)=
	\frac{1}{2}\,\left(\frac{p^K}{p^J}+\frac{p^J}{p^K}-\frac{p^J\,p^K}{(p^I)^2}\right)
\end{equation}
(I, J, K) an even permutation of 1, 2, 3; $p^I=\epsilon^{IKL}\,a_K\,a_L$\,.

Derive now the classical equations transforming to a new canonical variables $\pi_I$ and 
$q^I$ and to new time coordinate:
%( 23 ) 
\begin{equation}
\begin{array}{c}\ds
  \pi_I=-(\log {a_I})^{'}\,, \q q^I=\frac{1}{2}\,\log p^{I}\,, 
	\q dt=a_1\,a_2\,a_3\, d{\tau} \\[4mm]\ds
  \{ q^I,\,\pi_J\} = \kappa\,\delta_{J}^{I}\,, 
  \end{array} 
\end{equation}
 
Separating terms with momenta like variables $\pi_I$, we obtain the potential term
%( 24 ) 
\begin{equation}
	W(a_1,\,a_2,\,a_3)=\frac{1}{2}\,\left[\sum_I\,a_{I}^{4}\right]-a_{1}^{2}\,a_{2}^{2}-
	a_{2}^{3}\,a_{3}^{2}-a_{1}^{2}\,a_{3}^{2}\,.
\end{equation}
 
From (22) and (23) it is seen that at the small $a_I$ (or $p^I$) due to the 
divergencies of the spin connection components there is singularity.

Classical equation of motion are 
%( 25 )
\begin{equation}
	\frac{1}{2}\,(\log a_1)^{''}=(a_{2}^{2}-a_{3}^{2})^2-a_{1}^{4}
\end{equation}
and two e.o.m. by cyclic.

Right hand sides in (25) are vanished for Bianchi I geometry, giving 
$a_I \sim t^{\alpha_I}$ with, $\sum_{I}=1=\sum_{I}\,\alpha_I^2$, i.e. the 
Kasner solution.

For Bianchi IX geometry one may write the evolution potential in terms of $p^I$:
\begin{equation}
\begin{array}{c}\ds
	W(p^1,\,p^2,\,p^3)=2\,[(p^1\,p^2\,(\Gamma_1\,\Gamma_2-\Gamma_3)+p^3\,p^1\,
	(\Gamma_3\,\Gamma_1-\Gamma_2)\\[4mm]\ds
	+p^2\,p^3\,(\Gamma_2\,\Gamma_3-\Gamma_1)]
  \end{array} 
\end{equation}
which has an infinite walls at $p^I \approx 0$ due to the divergence of the 
spin connection components. Evolution consists of the succession of the Kasner 
epochs with reflections on the walls and this process never stops what leads to 
the BKL chaos.

To be closer to the common picture one can introduce the Misner variables 
$$
	\Omega=-\frac{1}{3}\,\log V=-\frac{1}{3}\, \log{(a_1\,a_2\,a_3)}
$$
and anisotropies $\beta_I$: 
$$
\begin{array}{c}\ds
  a_1=e^{-\Omega+\beta_++\sqrt{3}\,\beta_-}\,, \q  
  a_2=e^{-\Omega+\beta_++\sqrt{3}\,\beta_-}\,,\\[4mm]\ds
  a_3=e^{-\Omega-2\,\beta_+} 
  \end{array} 
$$
 
Then the potential (26) takes the form 
%( 27 )
\begin{equation}
\begin{array}{c}\ds
  W(\Omega,\,\beta_+,\,\beta_-)=\frac{1}{2}\,e^{-4\Omega}\,[e^{-8\,\beta_+}-4\,
  e^{-2\,\beta_+}\,\cosh(2\sqrt{3}\,\beta_-)\\[4mm]\ds
  2\,e^{4\,\beta_+}(\cosh(4\,\sqrt{3}\,\beta_-)-1)]\,.
  \end{array}
\end{equation}

Volume dependence factorizes and the rest anisotropy potential shows exponential 
walls for the large anisotropies. For instance, at 
typical wall has a form if one takes $\beta_- = 0$ and $\beta_+ < 0$: 
%( 28 )
\begin{equation}
	W \sim \frac{1}{2}\,e^{-4\,\Omega-8\,\beta_+}
\end{equation}

To prevent the "eternal" reflections at the walls where the expansion/contraction 
behavior of the different directions change, it is necessary to stop the unconstrained 
rise of the heights of the walls. Just the quantum effects are called up for this 
prevention.

In other words, they should lead to the upper limit on the curvature.

What is the concrete scenario to achieve this aim in the LQC?

First of all, one has to have in mind that in the LQG and in the L Q C there 
exist effective minimal length (or area, $A_{1/2} = 8\pi\,\beta\,l^2_p\,\sqrt{3/2}$)
or the maximal curvature.

Quantization according to the rules of the game in LQG results in the 
replacement of the spin connections components $\Gamma_I$ by the effective coefficients, 
which leads to the effective potential instead of (26).

Central moment here is the special rules of the quantization of the inverse densitized 
triad variables $(p^I)^{-1}$ or the inverse volume, giving that in the L Q G they are 
not singular at $p^I=0$ despite the classical curvature divergence.

One should remember that the LQC which actively considers various important 
phenomenological effects is not in the strong sense the direct limiting case of 
the full LQG where the desired boundedness of the inverse scale factors or the 
inverse volume are ensured. LQC is the usual cosmological mini superspace with 
its symmetry reduction and oversimplifications, quantized by the LQG methods and 
techniques. For this reason, for instance isotropic model in the LQC, in contrast 
to some claims, has no bounded from above inverse scale factor whereas in the full 
scale LQG it is proved that such a inverse scale and inverse volume operators have 
bounded from above eigenvalues [43]. The physical explanation of this not common 
situation may be that the isotropic homogeneous quantum fluctuations for isotropic 
mini superspace model are not enough to eliminate the classical singularity. For 
anisotropic Bianchi IX model it is not excluded that the fluctuations of the same 
symmetry may ensure this elimination although it is not based on the firm grounds as 
it has place for the LQG [43].

Anyway, taking the assumption that for Bianchi IX geometry this conjecture is realized,
let us continue to follow what is happened with the chaos near the singularity. We again 
follow [42]. For convenience, we take $\frac{1}{2}\,\beta\,l^2_p = 1$ making $p^I$ 
dimensionless.

Quantization replaces $(p^i)^{-1}$ in the spin connection components by the function 
$F(p/2j)$ where the parameter $j$ appears explicitly and controls the peak of the 
function $F$. The same parameter $j$ enters the expression of the area in the LQG:
$A(j) =8\,\pi\,\beta\,l^2_p\sqrt{j\,(j+1)}$. Further, follow the recipe of the 
LQG to obtain the effective description, one needs to replace all negative powers 
of triad $p^I$ with the appropriate factors of the spectrum of the inverse volume 
operator.

For instance, 
$$
p^{-3/2} \rightarrow d = D ( p / p^* )/ p^{3/2} 
$$
with $p^* = a^{*^2} = 16\,\pi\,j\,\mu_0\,/3$.
Function $D \approx 1$ for $p / p^* >> 1$, recovering the classical behavior.
For the small $p$ (or $a$), $D(\frac{p}{p^*}) \sim  p^{1/2}$ or $d \sim p^6$ thus 
giving the smooth behavior at the singularity.

For the anisotropic homogeneous model the components of $(p^I)^{-1}$ $(I=1,2,3)$ are 
replaced in the spin connection components by a function $F(p^I /2\, j)$ giving the 
effective spin connection. Parameter $j$ belongs to the set of the ambiguities of this 
framework and controls the peak of the function $F$. The resulting effective potential 
as a function of $p^I$ at fixed volume $V$ has a form
%( 29 ) 
\begin{equation}
\begin{array}{c}\ds
W_j\,\left[ p^1,\,p^1,\,\left(\frac{V}{p^1}\right)^2\right] = W_j(p^1,\,p^1,\,2jq) \approx
\\[4mm]\ds
\approx \frac{V^4\,F^2\,(q)}{32\,j^4\,q^2}\,[3-2q\,F\,(q)]
\end{array}
\end{equation}
where $q = \frac{1}{2j}\left(\frac{V}{p^i} \right)$.

At the peak and beyond it $F(q) \approx 1/q$ and we have the classical wall 
$\frac{1}{2}\,e^{-4\Omega-8\beta_+}$. 
The peak of the finite walls is reached for a constant $q$ which in the usual variables gives that 
$e^{-2\,\Omega+2\beta_+}= const$. Maxima of the wall lie on the line $\beta_+ = \Omega + const$ in 
the classical phase space and the height of the wall drops off as $e^{-12\,\Omega} \sim V^4$ with the 
decreasing $V \rightarrow 0$. 

At very small volume the walls collapse more rapidly and the effective potential becomes negative 
everywhere at the volumes close of in the Planck units. For the smallest value of $j = 1/2$ it 
is about Planck volume.

Thus, with the decreasing walls during the evolution towards the singularity the classical reflections 
will stop at a finite time interval and the chaos should disappear. Universe -at some time- can "jump 
over the wall" [42], Kasner regime becomes stable. If we think about non homogeneities, the patches of 
the corresponding space become of the order of a Planck volume, i.e. the scale of the discreteness. Below 
that scale further fragmentation does not happen and the discreteness is preserved. 

In the large volumes when the evolution is chaotic, two nearby points - patches will diverge away with 
no correlations between the points ("non interacting two dimensional gas"). In the vicinity of singularity, 
the chaotic motion is replaced by the Kasner evolution, points-patches begin to correlate ("interacting 
two dimensional gas").

The evolution to the singularity on the basis of LQC with its final non-chaotic scenario near the Planck 
scale and beyond rises the important question of the increasing the role of non-homogeneities at that 
scale. To sustain the homogeneous regime one needs the further and further fragmentation of patches of 
the spatial regions. But at the conditions of the discrete spatial geometry, the fragmentation must be 
stopped and the homogeneity should be replaced by non - homogeneity. Thus, at the approaching to the 
singularity role of the inhomogeneous quantum fluctuations may be essential. This brings us to the 
relatively old notion of the "turbulent" universe [50], [51].

\section{In lieu of conclusion}
 
In lieu of conclusion, we enumerate here several important problems considered by the 
LQC and not concerning the chaos in the cosmology.
LQC, using the similar approach (based on the minisuperspace and the recipes of the quantization from 
the full LQG) has a several important contribution to the "explanation" of the inflation [44], to the 
quantum nature of the Big Bang [45], possible observational signatures in the CMBR [46], avoidance of 
the future singularity [47] where the interesting effect of the negative quadratic density correction 
inspired by LQC is observed in the FRW equation, and the quantum evaporation of the naked singularity 
[48]. Last paper gives an interesting view on the problem of the gravitational collapse of the matter 
(scalar field as an example) near the classical singularity. The authors of [48] observed the rise of 
the strong outward energy flux which dissolves the collapsing cloud before the formation of the 
singularity. This effect based on the LQC may be considered as a mechanism of the censorship of the 
naked singularity [49].

Authors of [48] think about the observational signature of this effect in the astrophysical bursts.
Time will show how reliable are these interesting investigations.

\section*{Acknowledgments} 
%{\bf Acknowledgments}\\[1mm] 
It is my great pleasure to thank Remo Ruffini and Vahagn Gurzadyan for invitation to the 
superbly organized, in spite of hot weather, MG11 meeting.
I am grateful to Remo Ruffini for the support which made my participation possible.

\bibliographystyle{ws-procs975x65}
\bibliography{ws-pro-sample}

\end{document}